\documentclass{notpnastwo} 

\usepackage{graphicx}
\usepackage{amssymb,amsfonts,amsmath}

\newcommand{\expect}[1]{\left \langle #1 \right \rangle}                

\url{http://arxiv.org/abs/1105.2278}
\copyrightyear{2011}
\issuedate{Originally submitted on 11 May 2011}
\volume{Revised on 13 July 2011}
\issuenumber{arXiv:1105.2278}

\usepackage[colorlinks=true,citecolor=blue,linkcolor=blue]{hyperref}

\begin{document}

\title{Nonequilibrium candidate Monte Carlo:\\
A new tool for efficient equilibrium simulation}

\author{Jerome P. Nilmeier\affil{1}{Biosciences and Biotechnology Division, Physical and Life Sciences Directorate, Lawrence Livermore National Laboratory, Livermore, California 94550, USA},
Gavin E. Crooks\affil{2}{Physical Biosciences Division, Lawrence Berkeley National Laboratory, Berkeley, California 94720, USA},
David D. L. Minh\affil{3}{Biosciences Division, Argonne National Laboratory, Argonne, Illinois 60439, USA},
\and
John D. Chodera\thanks{To whom correspondence should be addressed. E-mail: \url{jchodera@berkeley.edu}}\affil{4}{California Institute for Quantitative Biosciences (QB3), University of California, Berkeley, California 94720, USA}}

%
%
%


\maketitle


\begin{article}
\begin{abstract}
Metropolis Monte Carlo simulation is a powerful tool for studying the equilibrium properties of matter.
In complex condensed-phase systems, however, it is difficult to design Monte Carlo moves with high acceptance probabilities that also rapidly sample uncorrelated configurations.
Here, we introduce a new class of moves based on \emph{nonequilibrium} dynamics: 
candidate configurations are generated through a finite-time process in which a system is actively driven out of equilibrium, and accepted with criteria that preserve the equilibrium distribution.
The acceptance rule is similar to the Metropolis acceptance probability, but related to the nonequilibrium work rather than the instantaneous energy difference.
Our method is applicable to sampling from both a single thermodynamic state or a mixture of thermodynamic states, and allows both coordinates and thermodynamic parameters to be driven in nonequilibrium proposals.
While generating finite-time switching trajectories incurs an additional cost, driving some degrees of freedom while allowing others to evolve naturally can lead to large enhancements in acceptance probabilities, greatly reducing structural correlation times.
Using nonequilibrium driven processes vastly expands the repertoire of useful Monte Carlo proposals in simulations of dense solvated systems.
\end{abstract}


\abbreviations{MC, Metropolis Monte Carlo; MD, molecular dynamics; MCMC, Markov chain Monte Carlo; NCMC, Nonequilibrium candidate Monte Carlo}


\dropcap{I}n this paper, we describe a new technique for constructing efficient Markov chain Monte Carlo (MCMC)~\cite{jun-s-liu:MCMC} moves that both have high acceptance rates and also allow rapid transit through configuration space, greatly enhancing convergence rates in simulations of dense solvated systems.
The Metropolis Monte Carlo~\cite{metropolis:jcp:1953:metropolis-monte-carlo, hastings:biometrika:1970:metropolis-hastings} sampling procedure is generalized by using \emph{nonequilibrium} processes to generate candidates for \emph{equilibrium} simulations.
Within this framework, moves that are efficient for an isolated part of a system but lead to near-universal rejection in standard Monte Carlo simulations of dense mixtures can be converted to nonequilibrium processes that generate candidates with higher acceptance probabilities.
In this new procedure, the acceptance criteria is related to the nonequilibrium work, rather than the potential energy difference used in traditional Monte Carlo moves.

Since their introduction in the mid-twentieth century, 
Metropolis Monte Carlo (MC)~\cite{metropolis:jcp:1953:metropolis-monte-carlo, hastings:biometrika:1970:metropolis-hastings} and molecular dynamics (MD)~\cite{rahman:phys-rev:1964:molecular-dynamics} simulations have become favored tools for sampling from complex multidimensional distributions, such as configurations of microscopic physical systems in thermodynamic ensembles.
However, these methods can produce highly correlated samples, leading to slow convergence of estimated expectations.
While MD requires the use of small timesteps for numerical stability and to approximate sampling from the desired distribution, MC simulations can, in principle, make use of non-local moves that accelerate mixing of the Markov chain.
Indeed, vast improvements in efficiency have been obtained by applying cleverly constructed move sets that exploit physical intuition about the system under study, such as cluster moves in Potts and Ising model simulations~\cite{swendsen-wang:prl:1987:swendsen-wang,wolff:prl:1989:cluster-moves}.

\begin{figure}[tbp]
\noindent
\resizebox{\columnwidth}{!}{\includegraphics{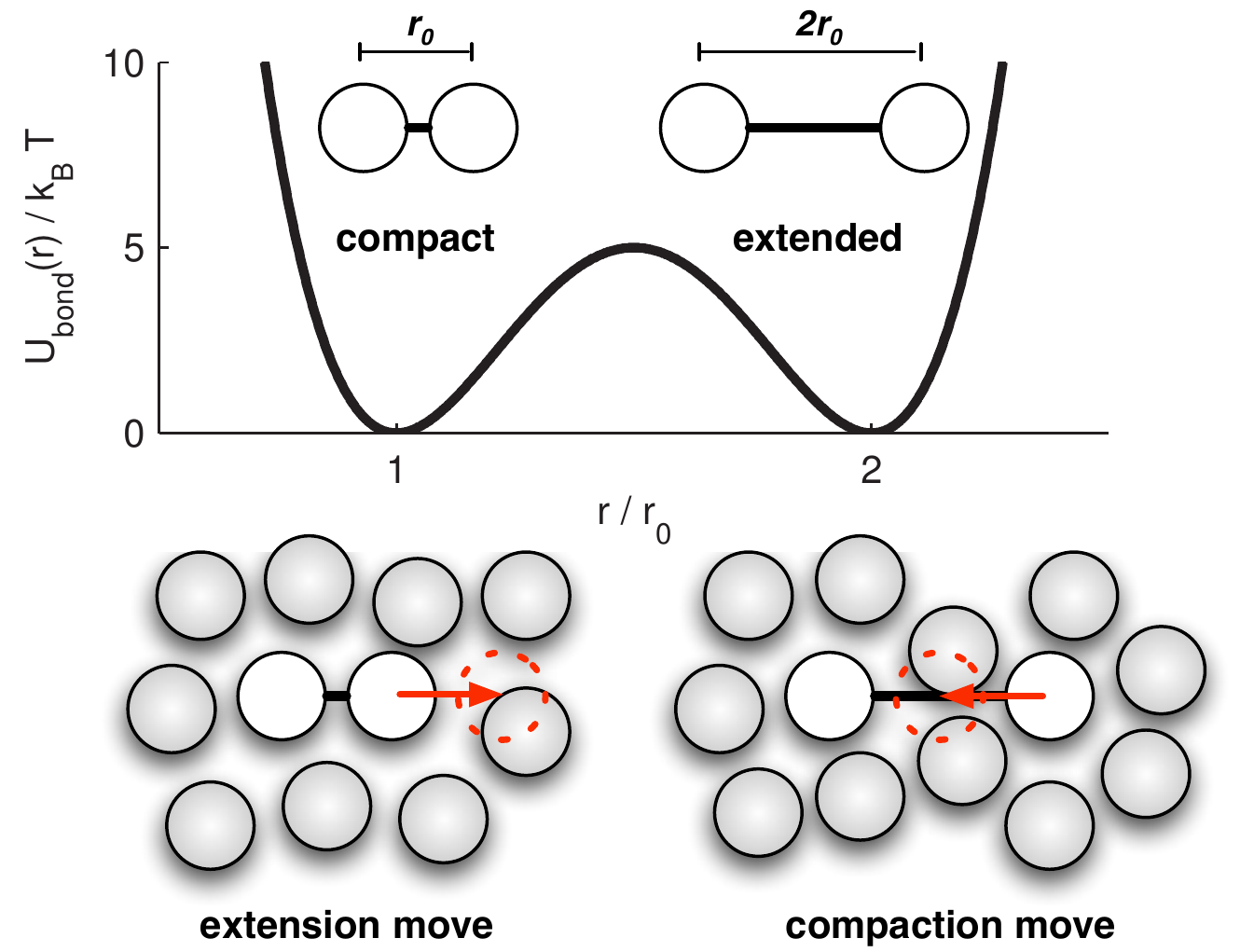}}
\caption{{\bf Bistable dimer potential and instantaneous MC moves in WCA fluid.}
An extension move increases the dimer extension by $\Delta r = +r_0$, while a compaction move decreases the dimer extension by $\Delta r = -r_0$. 
Both move types meet with near-universal rejection when implemented as instantaneous MC moves in a dense WCA fluid.
Note that the lower panel is only a cartoon --- the actual described simulation is of a 3D system.
\label{figure:wca-fluid-diagram}}
\end{figure}

Designing efficient moves requires striking a balance between rapid traversal of phase space and ensuring reasonable acceptance probabilities.
For complex heterogeneous systems such as solvated biomolecules, achieving this balance remains challenging.
Typically, efficient moves exploit physical insight into kinetically slow processes and energetically favorable configurations.
Often, the experimenter may possess physical insight about one component in the system (e.g.~a biomolecule) that permits the design of moves that would be efficient in the absence of other components (e.g.~solvent), but encounter energetically unfavorable interactions in their presence, reducing acceptance rates to levels where standard MC provides no benefit.
\color{black}
As an illustrative example, consider a bistable dimer---a pair of particles interacting with a potential with minima in compact or extended configurations, separated by a high barrier (see Fig.~\ref{figure:wca-fluid-diagram}).
For simulations of this system in a vacuum, a simple and effective standard MC move is to instantaneously increase the interparticle distance from a compact to extended configuration (or conversely, to decrease the distance from an extended to compact configuration).
When the dimer is immersed in a dense solvent, however, this move is met with near-universal rejection because solvent molecules overlap with proposed configurations.

\color{black}
One approach that can allow unperturbed degrees of freedom to relax, and hence maintain a reasonable acceptance rate, is to use a \emph{nonequilibrium} process to generate candidate configurations.
Using the appropriate acceptance criterion for the final configuration will preserve the equilibrium distribution.
\color{black}
In the case of the bistable dimer immersed in dense solvent, the extension (contraction) may be carried out over a finite number of increments interleaved with standard Metropolis Monte Carlo or molecular dynamics steps that allow the solvent to reorganize to avoid overlap with the dimer particles.

The basic idea of using nonequilibrium driven processes as Monte Carlo moves has precedents in both the statistical~\cite{neal:2004:tech-report:dragging-fast-variables, andrieu:jrssb:2010:particle-MC} and chemical~\cite{athenes:pre:2002:work-bias-mc, buergi:proteins:2002:constant-pH, stern:jcp:2007:constant-pH, nilmeier-jacobson:2009:jctc:posh} literature.  
Among the latter, Ath\`{e}nes developed ``work-bias Monte Carlo'' to enhance acceptance rates in grand canonical Monte Carlo simulations~\cite{athenes:pre:2002:work-bias-mc}, Stern presented a scheme to sample an equilibrium mixture of protonation states at constant pH in explicit solvent~\cite{stern:jcp:2007:constant-pH} (though an inexact variant was proposed earlier~\cite{buergi:proteins:2002:constant-pH}), and 
\color{black}
Nilmeier et al.~\cite{nilmeier-jacobson:2009:jctc:posh} proposed the driving of a subset of degrees of freedom to enhance acceptance rates (using an approximate acceptance criteria).
Nonequilibrium processes have also been used to generate configurations for parallel tempering simulations~\cite{opps-schofield:pre:2001:extended-state-space,brown-head-gordon:jcc:2003:cool-walk,jarzynski:pnas:2009:nonequilibrium-swaps}.

\color{black}
Here, we unify these ideas and significantly extend the application of nonequilibrium moves in physical simulations.
We present a theoretical framework, \emph{nonequilibrium candidate Monte Carlo} (NCMC), that is applicable to both single thermodynamic states (e.g.~NVT, NpT, $\mu$VT ensembles) as well as \emph{mixtures} of thermodynamic states (e.g.~expanded ensemble~\cite{lyubartsev:jcp:1992:expanded-ensembles,park:pre:2008:simulated-tempering} simulations).
Nonequilibrium proposals may drive a subset of degrees of freedom, the thermodynamic parameters characterizing the equilibrium distribution, or \emph{both}, significantly expanding the repertoire of Monte Carlo moves that lead to high acceptance and efficient mixing in dense condensed-phase systems.


\section{Equilibrium and Expanded Thermodynamic Ensembles}

For physical systems in equilibrium, the probability of observing a microstate is given by the Boltzmann distribution,
\begin{eqnarray}
\pi_\lambda(x) &=& Z_{\lambda}^{-1} \, e^{-u_\lambda(x)} \:\:;\:\: Z_{\lambda}=\int_\Gamma dx \, e^{-u_\lambda(x)}, \label{equation:fixed lambda_probability}
\end{eqnarray}
where $x \in \Gamma$ denotes a microstate of the system (which may include coordinates, momenta, and other dynamical variables, such as simulation box dimensions), $\lambda$ denotes a set of thermodynamic parameters whose values define a \emph{thermodynamic state}, and $Z_{\lambda}$ is a normalizing constant known as the \emph{partition function}.

The reduced potential $u_\lambda(x)$ depends on the thermodynamic ensemble under consideration~\cite{shirts-chodera:jcp:2008:mbar}.
For instance, in an isothermal-isobaric ($NpT$) ensemble, the reduced potential will assume the form,
\begin{eqnarray}
u_\lambda(x) &=& \beta [ H(x) + p V(x) ], \label{equation:reduced-potential}
\end{eqnarray}
which depends on the Hamiltonian $H(x)$ (which may include an external biasing potential, and is presumed to be invariant under momentum inversion) and the system volume $V(x)$.
In this ensemble, the vector of controllable thermodynamic parameters $\lambda \equiv \{\beta, H, p\}$ includes the inverse temperature $\beta$, the Hamiltonian $H(x)$, and external pressure $p$.  
Other thermodynamic parameters and their conjugate variables can be included or excluded to generate alternative physical (or unphysical) ensembles.

To allow sampling from \emph{multiple} thermodynamic states within a single simulation, we also define an \emph{expanded ensemble}~\cite{lyubartsev:jcp:1992:expanded-ensembles,park:pre:2008:simulated-tempering}, which specifies a joint distribution for $(x,\lambda)$ in a weighted mixture of thermodynamic states,
\begin{eqnarray}
\pi(x,\lambda) &=& \frac{Z_\lambda \pi_\lambda(x) \, \omega_\lambda}{\sum\limits_{\nu \in \mathcal{G}} \int_\Gamma dy \, Z_\nu \pi_\nu(y) \, \omega_\nu}, \label{equation:expanded-ensemble}
\end{eqnarray}
where $\omega_\lambda > 0$ specifies an externally-imposed weight for state $\lambda$.
Here, $\lambda \in \mathcal{G}$ may assume values in a discrete or continuous space $\mathcal{G}$.
If the set $\mathcal{G}$ consists of a single value of $\lambda$, a single thermodynamic state is sampled, and $\pi(x,\lambda) = \pi_\lambda(x)$. 
These thermodynamic states may correspond to a variety of different states of interest, such as temperatures in a simulated tempering simulation~\cite{marinari-parisi:europhys-lett:1992:simulated-tempering}, alchemical states in a simulated scaling simulation~\cite{yang:jcp:2007:simulated-scaling}, or protonation states in a constant-pH simulation~\cite{mongan:j-comput-chem:2004:constant-ph}.


\section{Nonequilibrium Candidate Monte Carlo}

We first describe the general form of NCMC.
At the start of an iteration, the current sample in the Markov chain, $(x^{(n)},\lambda^{(n)})$, which is assumed to be drawn from $\pi(x,\lambda)$, is used to initialize a trajectory, $(x_0, \lambda_0) = (x^{(n)},\lambda^{(n)})$.
A \emph{candidate} configuration $(x_T,\lambda_T)$ is then proposed through a nonequilibrium process in which a set of degrees of freedom and/or thermodynamic parameters may be driven according to some protocol~\cite{jarsynzki:j-stat-phys:2000:fluctuation-theorem} selected with a probability dependent only on $(x_0,\lambda_0)$.
Even if we only wish to sample from a single thermodynamic state $\lambda$, we may use a protocol that transiently drives the thermodynamic parameters away from $\lambda$ and back again (as in Ref.~\cite{brown-head-gordon:jcc:2003:cool-walk}).
Finally, an acceptance probability is computed and used to decide whether the next sample in the Markov chain, $(x^{(n+1)},\lambda^{(n+1)})$, is the candidate, $(x_T,\lambda_T)$, or the momentum reversal of the initial sample, $(\tilde{x}^{(n)},\lambda^{(n)})$.

An NCMC move begins by selecting a protocol $\Lambda$ from a set of possible protocols with probability $P(\Lambda | x_0, \lambda_0)$, such that there exists a reverse protocol labeled as $\tilde{\Lambda}$ (to be defined momentarily) with $P(\tilde{\Lambda} | \tilde{x}_T, \lambda_T) > 0$.
A protocol $\Lambda$ specifies both a series of $T$ \emph{perturbation kernels} $\alpha_t(x,y)$ and \emph{propagation kernels} $K_t(x,y)$, arranged in an alternating pattern such that $\Lambda \equiv \{ \alpha_1, K_1, \alpha_2, K_2, \ldots, \alpha_T, K_T \}$.
Both $\alpha_t(x,y)$ and $K_t(x,y)$ are conditional probabilities of $y \in \Gamma$ given any $x \in \Gamma$, and must satisfy the requirement that if $p(x,y) > 0$, then $p(y,x) > 0$, for $p$ substituted by $\alpha_t$ and $K_t$.

Each perturbation kernel $\alpha_t$ drives some or all of the degrees of freedom $x$ in a stochastic or deterministic way (e.g.~by driving a torsion angle, a distance between two atoms, or the volume of the simulation cell).
Similarly, each propagation kernel $K_t$ propagates some or all of the coordinates of the system at fixed $\lambda_t$ according to some form of MCMC or MD (e.g.~Metropolis Monte Carlo~\cite{metropolis:jcp:1953:metropolis-monte-carlo,hastings:biometrika:1970:metropolis-hastings}, velocity Verlet~\cite{swope:jcp:1982:velocity-verlet} deterministic dynamics, or overdamped Langevin stochastic dynamics~\cite{ermak-yeh:cpl:1974:brownian-dynamics,ermak:jcp:1975:brownian-dynamics}) that may also depend on the time index $t$.
Interleaving perturbation and propagation allows for energetically unfavorable interactions introduced by perturbation to be relaxed during propagation, potentially increasing acceptance rates relative to the instantaneous perturbations of standard Metropolis Monte Carlo.

The procedure by which a trajectory $X \equiv (x_0, x_1, \ldots, x_T)$ is generated from an initial microstate $x_0$ according to a protocol $\Lambda$ can be illustrated by the scheme,
\begin{eqnarray}
x_0 \stackrel{\alpha_1}{\longrightarrow} x_1^* \stackrel{K_1}{\longrightarrow} x_1 \rightarrow \cdots \longrightarrow x_{T-1} \stackrel{\alpha_T}{\longrightarrow} x_T^* \stackrel{K_T}{\longrightarrow} x_T
\end{eqnarray}
Application of the perturbation $\alpha_t$ to $x_{t-1}$ generates a perturbed configuration $x_t^*$, which is then propagated by the kernel $K_t$ to obtain $x_t$.

The reverse protocol $\tilde{\Lambda} \equiv \{K_T, \alpha_T, \ldots, K_0, \alpha_0\}$ reverses the order in which the perturbation and propagation steps are applied, generating the time-reversed trajectory $\tilde{X} \equiv \{\tilde{x}_T, \ldots, \tilde{x}_0\}$, where $\tilde{x}$ denotes $x$ with inverted momenta,
\begin{eqnarray}
\tilde{x}_T \stackrel{K_T}{\longrightarrow} \tilde{x}_T^* \stackrel{\alpha_T}{\longrightarrow} \tilde{x}_{T-1} \rightarrow \cdots \rightarrow \tilde{x}_1 \stackrel{K_1}{\longrightarrow} \tilde{x}_1^*  \stackrel{\alpha_1}{\longrightarrow} \tilde{x}_0 
\end{eqnarray}

The next step in NCMC is to accept or reject $(x_T,\lambda_T)$ as the next sample in the chain.  
To ensure that the stationary distribution $\pi(x,\lambda)$ is preserved, we enforce a strict \emph{pathwise} form of detailed balance,\footnote{\color{black}The described pathwise detailed balance condition is closely related to ``super-detailed balance'' (see, e.g.~\cite{frenkel:2004:pnas:rejected-states}), but also accounts for momentum reversal to extend its definition to include molecular dynamics integrators.}
\begin{eqnarray}
\lefteqn{A(X | \Lambda) \, \Pi(X | x_0, \Lambda) \, P(\Lambda | x_0, \lambda_0) \, \pi(x_0, \lambda_0)} \nonumber \\
&=& A(\tilde{X} | \tilde{\Lambda}) \, \Pi(\tilde{X} | \tilde{x}_T, \tilde{\Lambda}) \, P(\tilde{\Lambda} | \tilde{x}_T, \lambda_T) \, \pi(\tilde{x}_T, \lambda_T) . \label{equation:pathwise-detailed-balance}
\end{eqnarray}
The quantity $A(X | \Lambda)$ is the NCMC acceptance probability, while $\Pi(X | x_0, \Lambda)$ and $\Pi(\tilde{X} | \tilde{x}_T, \tilde{\Lambda})$ denote the probability of generating trajectory $X$ from initial configuration $x_0$ using protocol $\Lambda$, or $\tilde{X}$ from final configuration $\tilde{x}_T$ with protocol $\tilde{\Lambda}$, respectively,
\begin{eqnarray}
\Pi(X | x_0) &=& \prod_{1 \le t \le T} \alpha_t(x_{t-1}, x_t^*) \, K_t(x_t^*, x_t) \\
\Pi(\tilde{X} | \tilde{x}_T) &=& \prod_{T \ge t \ge 1} \alpha_t(\tilde{x}_t^*, \tilde{x}_{t-1}) \, K_t(\tilde{x}_t, \tilde{x}_t^*) .
\label{equation:forward-transition-probability}
\end{eqnarray}
{\color{black}Summation of Eq. \ref{equation:pathwise-detailed-balance} over all trajectories starting with $x_0$ and ending with $x_T$ recovers the standard detailed balance condition (see Appendix for proof).}

We define the ratio of proposal kernels as,
\begin{eqnarray}
\frac{\alpha(\tilde{X}|\tilde{\Lambda})}{\alpha(X|\Lambda)} &\equiv& \prod_{t=1}^{T} \frac{\alpha_t(\tilde{x}_t^*, \tilde{x}_{t-1})}{\alpha_t(x_{t-1}, x_t^*)} \label{equation:ratio-of-perturbation-kernels},
\end{eqnarray}
and the ratio of propagation kernels as the exponentiated difference in forward and backward conditional path actions as,
\begin{eqnarray}
e^{-\Delta \mathcal{S}(X|\Lambda)} &\equiv& \prod_{t=1}^T \frac{K_t(\tilde{x}_t, \tilde{x}_t^*)}{K_t(x_t^*, x_t)} . \label{equation:ratio-of-propagation-kernels}
\end{eqnarray}
Using the above expressions and the momentum invariance property $\pi(x,\lambda) = \pi(\tilde{x},\lambda)$, we may write the ratio of acceptance probabilities as,
\begin{eqnarray}
\lefteqn{\frac{A(X|\Lambda)}{A(\tilde{X}|\tilde{\Lambda})} =\frac{\pi(\tilde{x}_T, \lambda_T)}{\pi(x_0, \lambda_0)} \frac{P(\tilde{\Lambda} | \tilde{x}_T, \lambda_T)}{P(\Lambda|x_0,\lambda_0)} \frac{\Pi(\Tilde{X}|\tilde{x}_T, \tilde{\Lambda})}{\Pi(X | x_0, \Lambda)}} \nonumber \\
&=&\frac{\pi(x_T,\lambda_T)}{\pi(x_0,\lambda_0)} \frac{P(\tilde{\Lambda} | \tilde{x}_T, \lambda_T)}{P(\Lambda|x_0,\lambda_0)} \prod_{t=1}^{T} \frac{\alpha_t(\tilde{x}_t^*, \tilde{x}_{t-1})}{\alpha_t(x_{t-1}, x_t^*)} \frac{K_t(\tilde{x}_t, \tilde{x}_t^*)}{K_t(x_t^*, x_t)} \nonumber \\
&\equiv& \frac{\omega_T}{\omega_0} \frac{P(\tilde{\Lambda}|\tilde{x}_T,\lambda_T)}{P(\Lambda | x_0, \lambda_0)}  \frac{\alpha(\tilde{X}|\tilde{\Lambda})}{\alpha(X|\Lambda)} e^{-\Delta \mathcal{S}(X|\Lambda) - \Delta u(X|\Lambda)} 
\label{equation:acceptance-criteria-relationship}
\end{eqnarray}
where $\Delta u(X|\Lambda) \equiv u_T(x_T) - u_0(x_0)$ is the energy difference.
Eq.~\ref{equation:acceptance-criteria-relationship} is the main result of this paper, and is highly general with regard to both the choice of protocol for perturbation and propagation.
In subsequent sections, we discuss specific choices for these protocols that lead to particularly simple acceptance criteria.

Many choices of acceptance probabilities $A(X|\Lambda)$ that satisfy Eq.~\ref{equation:acceptance-criteria-relationship} are possible, including the well-known Metropolis-Hastings criterion~\cite{metropolis:jcp:1953:metropolis-monte-carlo,hastings:biometrika:1970:metropolis-hastings},
\begin{eqnarray}
\lefteqn{A(X|\Lambda) =} \label{equation:modified-metropolis-criteria} \\
&&\min \left\{ 1, \frac{\omega_T}{\omega_0} \frac{P(\tilde{\Lambda}|\tilde{x}_T,\lambda_T)}{P(\Lambda | x_0, \lambda_0)}  \frac{\alpha(\tilde{X}|\tilde{\Lambda})}{\alpha(X|\Lambda)} e^{-\Delta \mathcal{S}(X|\Lambda) - \Delta u(X|\Lambda)}  \right\} .  \nonumber
\end{eqnarray}
After generating $(x_T,\lambda_T)$ and evaluating $A(X|\Lambda)$, we generate a uniform random variate $U$.  If $U < A(X|\Lambda)$, then the candidate becomes the next value in the chain, $(x^{(n+1)},\lambda^{(n+1)}) = (x_T,\lambda_T)$.  Otherwise, it  is rejected, we perform a momentum flip, and the next value becomes $(x^{(n+1)},\lambda^{(n+1)})=(\tilde{x}_0,\lambda_0)$.
Alternately, we may perform a momentum flip upon acceptance, $(x^{(n+1)},\lambda^{(n+1)}) = (\tilde{x}_T,\lambda_T)$ and preserve the momentum upon rejection, $(x^{(n+1)},\lambda^{(n+1)})=(x_0,\lambda_0)$.
We cannot, however, ignore the momentum flip completely; as explained in the Appendix, it is necessary to preserve the equilibrium distribution.

We note that NCMC need not be used exclusively to sample from $\pi(x,\lambda)$, but can be mixed with other MCMC moves or with MD~\cite{jun-s-liu:MCMC}.
For example, one may reinitialize velocities from the Maxwell-Boltzmann distribution after each NCMC step; this is a Gibbs sampling MCMC move using the marginal distribution for velocities.


\section{Perturbation Kernels}

A large variety of choices are available for the perturbation kernels $\alpha_t(x,y)$.
Through judicious selection of these kernels, a practitioner can design nonequilibrium proposals that carry some component of the system from one high-probability region to another with high acceptance rates.
We briefly describe a few possibilities.

\subsection{Stochastically Driven Degrees of Freedom}
Suppose we wish to drive a  torsion angle $\phi$ (an angle subtended by four bonded atoms) \emph{stochastically} by rotating it to a new torsion angle $\phi'$ (holding bond lengths and angles fixed)according to some probability, such as the von Mises circular distribution centered on $\phi$,
\begin{equation}
\eta(\phi' | \phi) = [2 \pi I_0(\kappa)]^{-1} e^{\kappa \cos(\phi' - \phi)} ,
\end{equation}
with $I_0(\kappa)$ denoting the modified Bessel function of order zero and $\kappa > 0$ taking the role of a dimensionless force constant.
Because the stochastic perturbation is made in a non-Cartesian coordinate, a Jacobian $J(\phi)$ must be included to compute $\alpha(x,y)$ in Cartesian coordinates, resulting in the ratio,
\begin{equation}
\frac{\alpha_t(\tilde{y},\tilde{x})}{\alpha_t(x,y)} = \frac{\eta(\phi|\phi') J(\phi)}{\eta(\phi'|\phi) J(\phi')} = 1.
\end{equation}
where $J(\phi') = J(\phi) =1$ because the transformation (a rotation about a bond vector) preserves the Cartesian phase space volume.

\subsection{Deterministically Driven Degrees of Freedom}

Instead of perturbing the torsion angle stochastically, we can deterministically drive it in small, fixed increments $\Delta \phi$.
In this case, we effectively define an invertible map $\mathcal{M}$ that takes $x \rightarrow y$, such that $y = \mathcal{M} x$ differs from $x$ only in the rotation of the specified torsion $\phi$ by $\Delta \phi$.
To implement this, we may choose a perturbation $\Delta \phi$ from a distribution where $\pm \Delta \phi$ have equal probability, and drive $\phi(x)$ from its current value $\phi_0$ to a final value $\phi_T = \phi_0 + \Delta \phi$ over $T$ steps in equal increments, such that $\phi(x_t)$ is constrained to $\phi_t \equiv (1-t/T) \phi_0 + (t/T) \phi_T$.
In this case, $\alpha_t(x,y) = \delta(y - \mathcal{M} x) J(x)$, where the Jacobian $J(x)$ represents the factor by which Cartesian phase space is compressed on the application of the map $\mathcal{M}$, which is again unity for rotation about a torsion angle by $\Delta \phi$, and, due to the invertibility of the map, the ratio $\alpha_t(\tilde{y}, \tilde{x}) / \alpha_t(x,y) = 1$.

\subsection{Simulation Box Scaling}

Another possible deterministic perturbation kernel is simulation box scaling.
A barostat can be implemented by proposing propagation kernels that scale the molecular centers and box geometry by a factor $s = [(V(x)+\Delta V)/V(x)]^{1/3}$ with $\Delta V$ chosen uniformly from $[V-\Delta V_0,V+\Delta V_0]$ applied as a factor of $s^{1/T}$ over the course of $T$ steps.
In this case, the perturbation kernel $\alpha_t(x,y)$ is a delta function that compresses or expands the molecular centers and box geometry.
Since the Jacobian is the ratio of infinitesimal volumes upon scaling, the ratio of perturbation kernels is $\alpha(\tilde{X}|\tilde{\Lambda}) / \alpha(X|\Lambda) = s^{3 N}$, where $N$ denotes the number of molecular centers.

\subsection{Thermodynamic Perturbation}

In many driven nonequilibrium processes, there is no direct perturbation to the coordinates, such that $\alpha_t(x,y) = \delta(x-y)$ and the ratio $\alpha(\tilde{X}|\tilde{\Lambda}) / \alpha(X|\Lambda) = 1$.
Instead, only the thermodynamic parameters $\lambda$ are varied in time, carrying the system out of equilibrium through action of the $K_t$ propagation kernels. 
We recover Neal's method~\cite{neal:2004:tech-report:dragging-fast-variables} if the reduced potential $u_t$ is a simple linear interpolation such that $u_t(x) = (1-t/T) u_0(x) + (t/T) u_T(x)$, the probability of choosing protocol $\Lambda$ is symmetric with $\tilde{\Lambda}$, and MC~\cite{metropolis:jcp:1953:metropolis-monte-carlo, hastings:biometrika:1970:metropolis-hastings} is used for the propagation kernel $K_t$.


\section{Propagation Kernels}

The choice of propagation kernels available is also very broad.
If strong driving is performed in selection of $\alpha$, one may elect to choose a time-independent propagation kernel $K_t(x,y) \equiv K(x,y)$ that samples from a stationary distribution $\pi(x)$ of interest.
Alternatively, a strongly time-dependent $K_t$ could be selected to transiently drive the system out of equilibrium, or from the equilibrium distribution at one thermodynamic state to another.
Some possible choices are described below.


\subsection{Reversible Markov Chain Monte Carlo}
One may propagate some or all of a system's degrees of freedom (e.g. those not affected by the perturbation kernel $\alpha_t$) by a method that satisfies detailed balance in $\pi_t$,
\begin{eqnarray}
\pi_t(x) K_t(x,y) &=& \pi_t(\tilde{y}) K_t(\tilde{y},\tilde{x}) ,
\label{eq:detailed-balance}
\end{eqnarray}
where $\pi_t(x) \equiv Z_t^{-1} e^{-u_t(x)}$ for a specified $u_t(x)$.
Many MCMC methods~\cite{jun-s-liu:MCMC}, including Metropolis~\cite{metropolis:jcp:1953:metropolis-monte-carlo, hastings:biometrika:1970:metropolis-hastings} and various hybrid Monte Carlo (HMC) algorithms that combine discrete-timestep integrators with Monte Carlo acceptance/rejection steps~\cite{duane:1987:phys-lett-b:hybrid-monte-carlo,lelievre-stoltz-rousset:2010:GHMC}, obey detailed balance and are commonly used for the simulation of physical systems.

By analogy with Crooks~\cite{crooks:jsp:1998:neq}, we define a \emph{work} $w$ and \emph{heat} $q$ for the nonequilibrium driven process,
\begin{eqnarray}
w(X|\Lambda) & = & \sum_{t=1}^T \left[ u_t(x_t^*) - u_{t-1}(x_{t-1}) \right] \label{eq:work} \\
q(X|\Lambda) & = & \sum_{t=1}^T \left[ u_t(x_t) - u_t(x_t^*) \right] \label{eq:heat}
\end{eqnarray}
such that $w(X|\Lambda)+q(X|\Lambda) = \Delta u(X|\Lambda)$, a restatement of the first law of thermodynamics.

The conditional path action difference can then be written in terms of the heat of the process, $q(X|\Lambda)$,
\begin{eqnarray}
\Delta \mathcal{S}(X | \Lambda)
= -\ln \prod_{t=1}^{T} \frac{\pi_t(x_t^*)}{\pi_t(x_t)}
= -q(X|\Lambda),
\end{eqnarray}
leading to an acceptance probability similar to standard MC, except that the work, $w(X|\Lambda)$, replaces the instantaneous potential energy difference,
\begin{eqnarray}
\frac{A(X|\Lambda)}{A(\tilde{X}|\tilde{\Lambda})} &=& \frac{\omega_T}{\omega_0} \frac{P(\tilde{\Lambda}|\tilde{x}_T,\lambda_T)}{P(\Lambda | x_0, \lambda_0)}  \frac{\alpha(\tilde{X}|\tilde{\Lambda})}{\alpha(X|\Lambda)} e^{-w(X|\Lambda)}.
\label{equation:detailed-balance-acceptance-criteria}
\end{eqnarray}


\subsection{Deterministic Dynamics}

When an isolated system is propagated by a symplectic integrator---a reversible, deterministic integrator that preserves phase space volume---the propagation kernels follow $K_t(x,y) = K_t(\tilde{y},\tilde{x})$.  
Hence, $\Delta \mathcal{S}(X|\Lambda) = 0$ and the acceptance ratio is,
\begin{eqnarray}
\frac{A(X|\Lambda)}{A(\tilde{X}|\tilde{\Lambda})} &=& \frac{\omega_T}{\omega_0} \frac{P(\tilde{\Lambda}|\tilde{x}_T,\lambda_T)}{P(\Lambda | x_0, \lambda_0)}  \frac{\alpha(\tilde{X}|\tilde{\Lambda})}{\alpha(X|\Lambda)} e^{-\Delta u(X|\Lambda)}
\label{equation:deterministic-acceptance-criteria},
\end{eqnarray}
where $\Delta u(X | \Lambda) \equiv u_T(x_T) - u_0(x_0)$ is the energy difference.
The equivalence of the work and energy difference for volume-preserving integrators was previously recognized in the context of fluctuation theorem calculations~\cite{jarzynski:prl:1997:neq-fe, lechner:2006:jcp:fast-switching}.

Symplectic integrators include velocity Verlet~\cite{swope:jcp:1982:velocity-verlet}.
These integrators are also symplectic when utilizing constraints through the application of algorithms such as RATTLE~\cite{andersen:1983:j-comput-phys:RATTLE}, provided that the constraints are iterated to convergence each timestep~\cite{leimkuhler-skeel:1994:j-comput-phys:symplectic-integrators-in-constrained-systems}.


\subsection{Stochastic Dynamics}
\color{black}
Stochastic integrators such as velocity Verlet discretizations of Langevin dynamics~\cite{paterlini-ferguson:chem-phys:1998:velocity-verlet-langevin,izaguirre:2010:psb:normal-mode-langevin} sample a modified distribution that differs from the desired equilibrium distribution $\pi_t$ in a timestep-dependent manner~\cite{lelievre-stoltz-rousset:2010:free-energy-computations}.
While this modified distribution may be difficult or impossible to compute in order to recover equilibrium properties by reweighting, computation of the relative action $\Delta \mathcal{S}(X|\Lambda)$ is relatively straightforward, and the NCMC acceptance criteria ensures that the NCMC-sampled configurations are distributed according to the desired equilibrium ensemble\footnote{{\color{black}Alternatives to using NCMC to correct stochastic integration include introducing a Metropolization correction after each step, as in the generalized hybrid Monte Carlo (GHMC) integrator we use in the example~\cite{gustafson:statistics-and-computing:1998:GHMC,jun-s-liu:MCMC,lelievre-stoltz-rousset:2010:free-energy-computations,lelievre-stoltz-rousset:2010:GHMC}.}}.
\color{black}
As examples, we compute $\Delta \mathcal{S}(X | \Lambda)$ for the overdamped Langevin (Brownian) dynamics integrator of Ermak and Yeh~\cite{ermak-yeh:cpl:1974:brownian-dynamics,ermak:jcp:1975:brownian-dynamics} and the Br\"{u}nger-Brooks-Karplus (BBK) Langevin integrator~\cite{brunger-brooks-karplus:cpl:1984:bbk-integrator,pastor-brooks-szabo:mol-phys:1988:bbk-integrator,schlick} in the Appendix.


\section{Illustrative Application: Bistable Dimer in a WCA Fluid}
\label{section:wca-application}

\begin{figure}[tbp]
\noindent
\resizebox{\columnwidth}{!}{\includegraphics{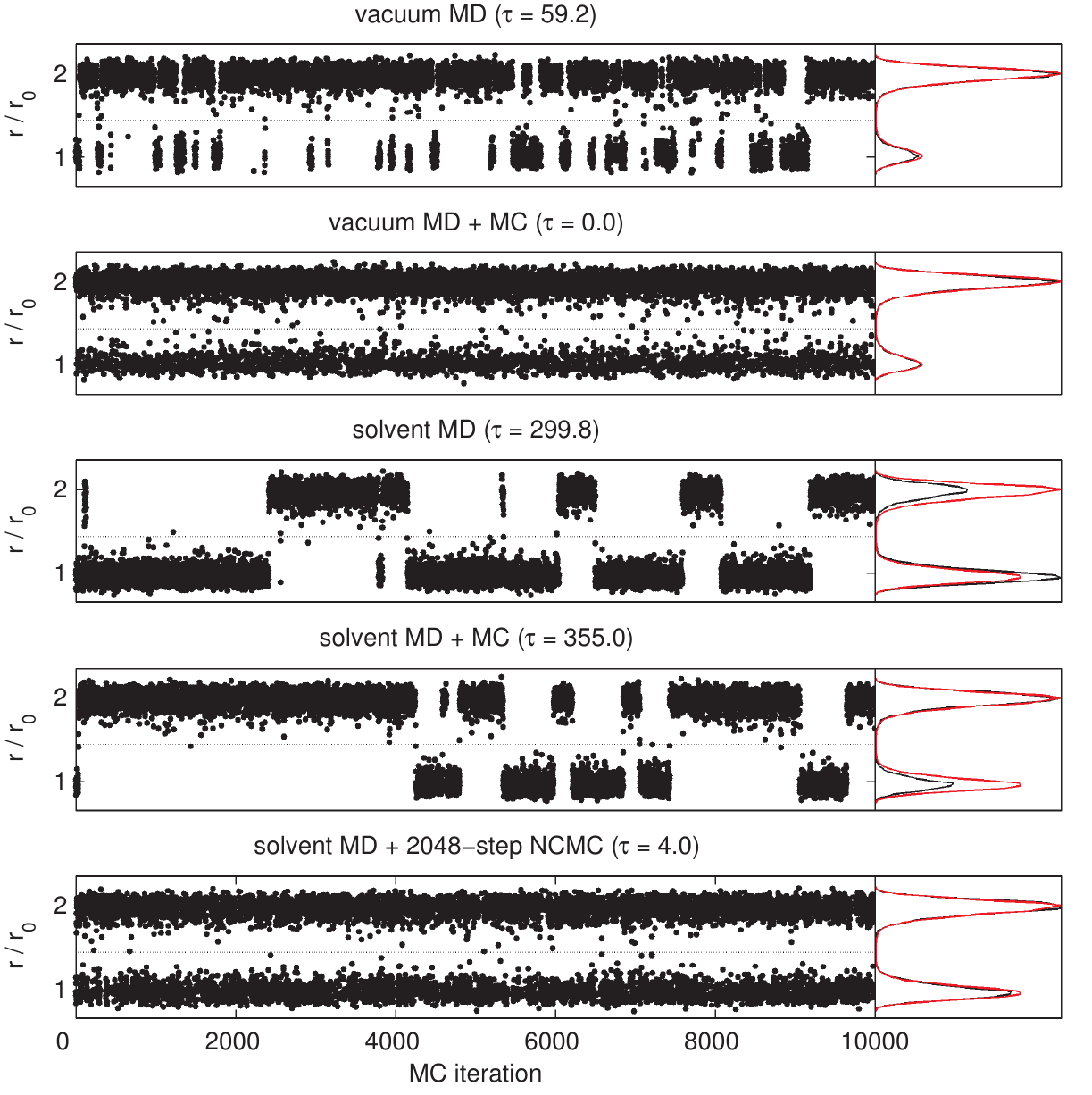}}
\caption{{\bf Trajectories of WCA dimer system in vacuum and solvent.}
\emph{Left:} The dimer extension $r$ as a function of simulation iteration.
The dotted horizontal line denotes division between compact and extended configurations.
The quantity $\tau$ printed above each plot indicates the estimated integrated autocorrelation time for the dimer extension $r$.
\emph{Right:} Histogram accumulated over trajectory (black), with true equilibrium distribution (red).
Plot titles denote whether simulation was run in vacuum (\emph{vacuum}) or dense WCA fluid (\emph{solvent}), and whether the simulation utilized only 500 GHMC steps per iteration (\emph{MD}) or included instantaneous MC (\emph{MC}) or 2048-step NCMC moves (\emph{NCMC}) following each iteration.
\label{figure:wca-trajectory}}
\end{figure}

To demonstrate NCMC, we ran simulations of a bistable dimer (adapted from Section 1.3.2.4 of Ref.~\cite{lelievre-stoltz-rousset:2010:free-energy-computations}) in vacuum as well as a dense fluid.
The dimer consists of a pair of ``bonded'' particles interacting via a double-well potential, with minima at $r = r_0$ (compact) and $r = 2 r_0$ (extended), and a $5 \, k_\mathrm{B} T$ barrier {\color{black}(see Fig.~\ref{figure:wca-fluid-diagram})}.
In the solvated simulations, the dimer was immersed in a dense bath (reduced density $\rho \sigma^3 = 0.96$) of particles that interact with the bonded particles and each other via the Weeks-Chandler-Andersen (WCA) soft repulsive potential~\cite{weeks-chandler-andersen:1971:jcp:wca-fluid}.
Each simulation ``iteration'' consisted of velocity reassignment from the Maxwell-Boltzmann distribution, 500 steps of generalized hybrid Monte Carlo (GHMC) dynamics~\cite{gustafson:statistics-and-computing:1998:GHMC,jun-s-liu:MCMC,lelievre-stoltz-rousset:2010:free-energy-computations,lelievre-stoltz-rousset:2010:GHMC} (essentially, a Metropolis-corrected form of Langevin dynamics, henceforth referred to here as MD), optionally followed by either an instantaneous MC move or an NCMC move.

\color{black}
The rate at which effectively uncorrelated samples are generated can be quantified in terms of the correlation time $\tau$ for the dimer extension $r(t)$ (shown in Fig.~\ref{figure:wca-trajectory}).
This time represents the asymptotic decay time constant for the correlation function $C(t) = \expect{r(0)r(t)}$, which will behave like
\begin{eqnarray}
C(t) &=& C_\infty + (C_0 - C_\infty) e^{-t/\tau}
\end{eqnarray}
for large $t$, where $C_0 = \expect{r^2}$ and $C_\infty = \expect{r}^2$.
The correlation time is related to the statistical inefficiency, $g = 1 + 2\tau$, a factor that describes the number of iterations necessary to generate an effectively uncorrelated sample~\cite{chodera:jctc:2007:ptwham}.

For the MD simulation in vacuum (Fig.~\ref{figure:wca-trajectory}, top trace), we observe slow hopping between compact and extended configurations with a correlation time $\tau = 59.2$ iterations, resulting in slow convergence of the histogram.  
Introducing instantaneous MC dimer extension/contraction moves that modify the dimer extension by $\Delta r = \pm r_0$ reduces the correlation time to $\tau \approx 0.0$, such that an uncorrelated configuration is generated after each iteration of 500 MD steps and one instantaneous MC step (Fig.~\ref{figure:wca-trajectory}, second trace from top).

\begin{figure}[tbp]
\noindent
\resizebox{\columnwidth}{!}{\includegraphics{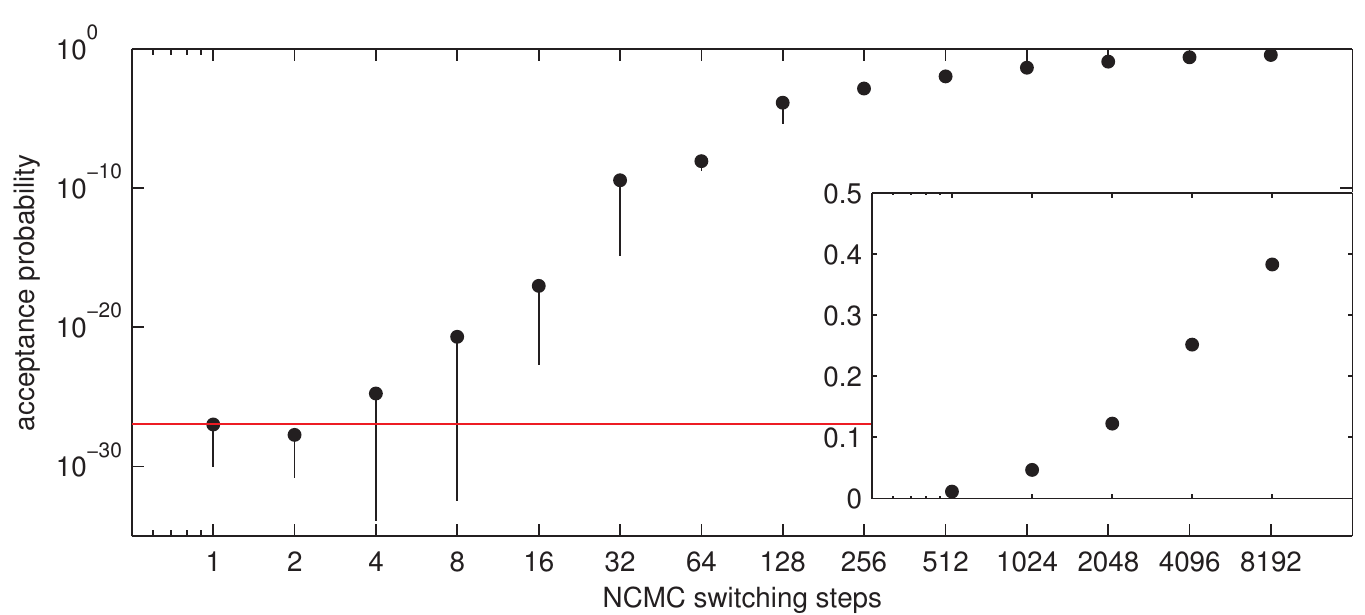}}
\caption{{\bf Acceptance probabilities of NCMC proposals.}
\color{black}
\emph{Top:} Acceptance probability of NCMC proposals as a function of length of nonequilibrium proposal trajectory (black dots), compared with instantaneous MC proposal (red line).
\emph{Inset:} Enlarged region with acceptance probability shown on linear scale.
Estimated 95\% confidence intervals are shown as vertical lines.
\color{black}
\label{figure:wca-acceptance}}
\end{figure}

When the dimer is immersed in a dense fluid of WCA particles, however, iterations consisting of 500 MD steps alone result in extremely slow barrier crossings, requiring g $\approx$ 600 iterations to produce an uncorrelated sample (Fig.~\ref{figure:wca-trajectory}, middle trace).
Unlike in vacuum, the introduction of instantaneous MC moves does not significantly reduce the correlation time $\tau$ (Fig.~\ref{figure:wca-trajectory}, second trace from bottom).
However, performing these same dimer expansion and contraction moves over 2048-step NCMC moves (Fig.~\ref{figure:wca-trajectory}, bottom trace) allows the system to rapidly mix between both compact and extended states with a correlation time of $\tau = 4.0$ iterations.
While each iteration requires a 5-fold increase in computational effort (500 MD steps + 2048 NCMC switching steps = 2548 force evaluations, versus 500 force evaluations for MD alone), a 67-fold reduction in correlation time is achieved, resulting in a remarkable order-of-magnitude gain in overall efficiency.

The length of the NCMC switching process is a free parameter that may be tuned to further improve efficiency.  Towards this end, we estimated the acceptance probability of the extension/contraction moves in dense solvent as a function of switching length (Fig.~\ref{figure:wca-acceptance}).
While instantaneous MC proposals of $\pm r_0$ are only accepted with probability $\approx 10^{-27}$ (the error is this quantity is likely underestimated due to its extremity),
dividing the move into smaller steps boosts the acceptance rate to a level useful in condensed-phase simulation.
If the move is divided into a small number of steps (1 to 8), there is little to no increase in acceptance rate, but for an intermediate number of steps (16 to 1024), there is a superlinear boost in the acceptance probability relative to the length of the switching process.
The acceptance probability finally reaches useful levels around 2000 steps, achieving an acceptance rate of 12\% using nonequilibrium proposal trajectories of 2048 steps, or 38\% for 8192 steps.

In general, there is no direct relationship between acceptance probability and efficiency.  Under certain assumptions relevant to the bistable dimer, however, it is possible to link the NCMC acceptance probability to $\tau_\mathrm{eff}$, an indirect estimate of the correlation time,
\begin{eqnarray}
\tau_\mathrm{eff} = \tau_\mathrm{MD} \left[ \frac{\tau_\mathrm{NCMC}}{\tau_\mathrm{MD} + \tau_\mathrm{NCMC}} \right],
\end{eqnarray}
where $\tau_\mathrm{MD}$ and $\tau_\mathrm{NCMC}$ are correlation times for iterations consisting solely of MD or NCMC moves, respectively.  The latter correlation time may be estimated from the average acceptance probability $\gamma$ by $\tau_\mathrm{NCMC} \approx -1/\ln (1-2 \gamma)$
(see \emph{Appendix} for derivation).

As shown in Fig.~\ref{figure:wca-efficiency}, the effective correlation time $\tau_\mathrm{eff}$ is only diminished when the NCMC acceptance probability is large enough such that $\tau_\mathrm{NCMC} \approx \tau_\mathrm{MD}$, which occurs when $\gamma \ge 0.13\%$ (about 256 switching steps or more).
For shorter switching times, even though the acceptance probability is high relative to instantaneous MC, it is still too small to significantly reduce $\tau_\mathrm{eff}$.

\begin{figure}[tbp]
\noindent
\resizebox{\columnwidth}{!}{\includegraphics{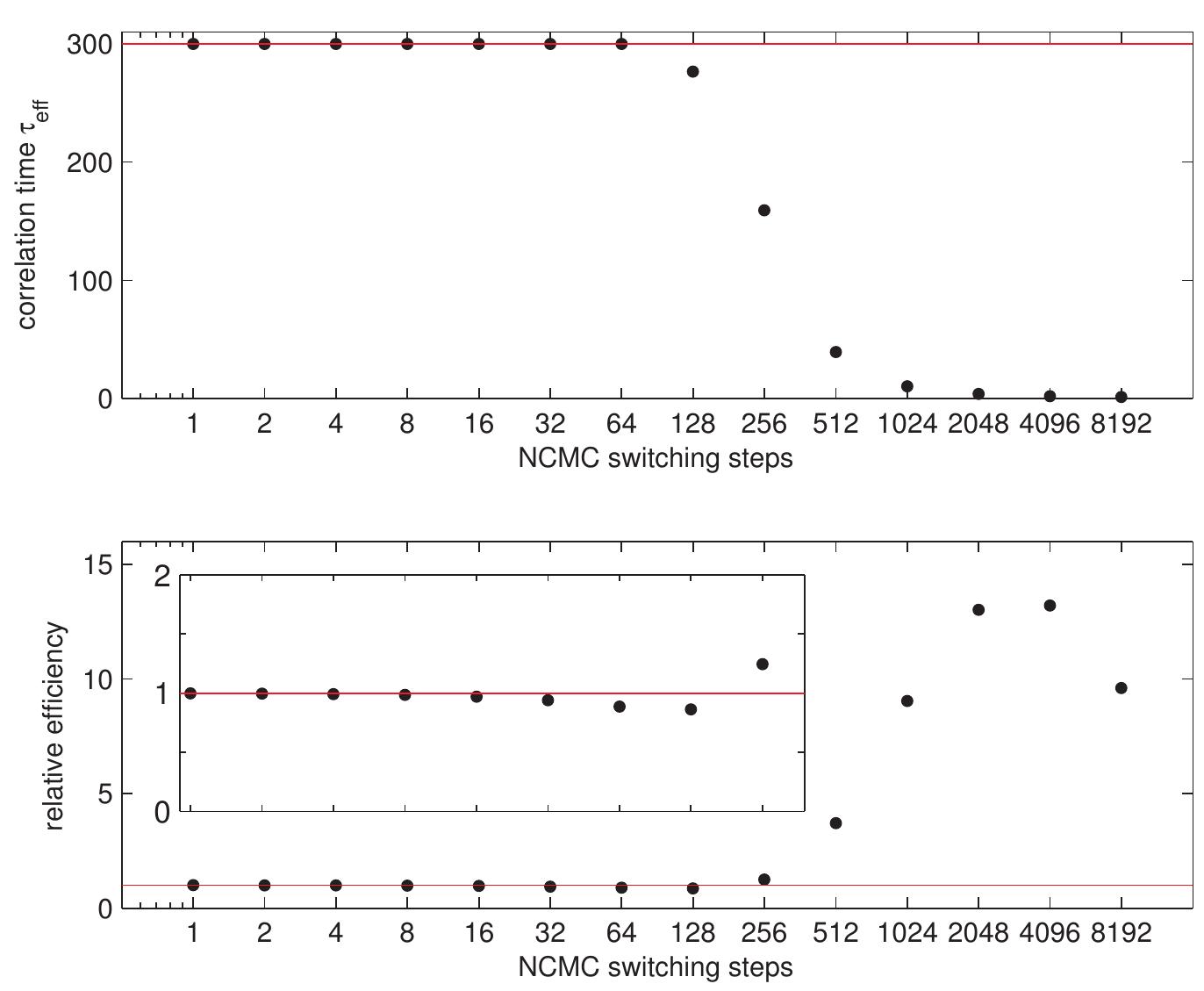}}
\caption{{\bf Statistical efficiency gain of NCMC proposals relative to instantaneous MC proposals.}
\color{black}
\emph{Top:} Effective correlation time $\tau_\mathrm{eff}$, in iterations, for MD+NCMC (black dots) compared to MD alone (red line).
\emph{Bottom:} Relative statistical efficiency of MD+NCMC, in terms of number of uncorrelated configurations generated for a fixed amount of computational effort, for MD+NCMC (black dots) relative to MD alone (red line).
\color{black}
\label{figure:wca-efficiency}}
\end{figure}

When comparing efficiency, we are most interested in the rate of generating uncorrelated configurations for a given amount of computational effort.
Relative to MD alone, this rate is described by the efficiency gain,
\begin{eqnarray}
E &\equiv& \frac{g_\mathrm{MD} T_\mathrm{MD}}{g_\mathrm{NCMC} (T_\mathrm{MD} + T_\mathrm{NCMC})},
\end{eqnarray}
Here,  $T_\mathrm{MD} = 500$ steps per iteration, and $T_\mathrm{NCMC}$ is again varied from 1 to 8192 steps.
The results are shown in the bottom panel of Fig.~\ref{figure:wca-efficiency}.
Surprisingly, there is actually a slight \emph{loss} in efficiency at short switching times---dropping to a minimum of $86.9\%$ the efficiency of MD alone at 128 steps---followed by a rapid gain in efficiency, plateauing at an efficiency gain of $\sim 13\times$ the efficiency of MD alone for 2048--4096-step NCMC proposals.
(A similar plateau behavior is observed in the modified parallel tempering protocol of Ref.~\cite{jarzynski:pnas:2009:nonequilibrium-swaps}.)
After this point, longer switching times do not achieve as high of an efficiency gain; even though the acceptance rate continues to increase as the number of NCMC switching steps is doubled again to 8192 steps, the reduction in correlation time is not sufficient to offset the additional cost of these moves.
\color{black}


\section{Epilogue}


\color{black}
We have described a procedure---\emph{nonequilibrium candidate Monte Carlo} (NCMC)---by which nonequilibrium proposals can be used within MCMC simulations to enhance acceptance rates.
In our illustration, we have demonstrated how its use can lead to large improvements in statistical efficiency---the rate at which uncorrelated samples are generated for a fixed amount of computational effort.
In other applications, whether similarly large efficiency gains are achieved will depend on the precise nature of the system under study and the nonequilibrium proposals introduced.
The most straightforward approach---borrowing Metropolis Monte Carlo proposals that are reasonable for one component of the system in isolation, and converting them to nonequilibrium proposals---is likely to be a fruitful avenue for generating efficient schemes, as was demonstrated here for a simple system.

More generally, the problem of selecting efficient nonequilibrium proposals is similar to the problem of choosing good reaction coordinates, in that it is desirable to drive the system along (possibly complex) slow collective coordinates where orthogonal degrees of freedom relax quickly.
The search for such collective coordinates is a topic of active research~\cite{bolhuis-chandler-dellago-geissler:annu-rev-phys-chem:2002:transition-path-sampling,best-hummer:pnas:2005:reaction-coordinates,ma-dinner:jpcb:2005:reaction-coordinates,e-ren-vanden-eijnden:jpcb:2005:finite-temperature-string,berezhkovskii-szabo:jcp:2005:reaction-coordinates,peters-trout:jcp:2006:reaction-coordinates,ensing:acc-chem-res:2006:metadynamics-reaction-coordinates}.
Given an initial nonequilibrium protocol, the issue of \emph{optimizing} such a protocol to minimize dissipation (and maximize acceptance) is also a topic of active study, led by forays into the world of single-molecule measurement~\cite{schmiedl-seifert:prl:2007:optimal-protocols,then-engel:pre:2008:optimal-protocol,seifert:jcp:2008:optimal-protocol}.
Recent work has also suggested that estimating the thermodynamic metric tensor along the nonequilibrium parameter switching path~\cite{salamon-berry:prl:1983:thermodynamic-length,crooks:2007:prl:thermodynamic-length,feng-crooks:2009:pre:thermodynamic-length,minh-chodera:2011:jcp:mis-estimator}, could prove useful in adaptively optimizing the switching protocol~\cite{shenfeld-xu-eastwood-dror-shaw:2009:pre:optimizing-thermodynamic-length}.

\color{black}
We note that switching trajectories contain potentially useful information.
Indeed, several methods~\cite{minh-adib:2008:prl:minh-adib,minh-chodera:2009:jcp:minh-chodera,minh-chodera:2011:jcp:mis-estimator} have recently been developed to estimate \emph{equilibrium} properties from nonequilibrium samples through the application of statistical estimator theory to nonequilibrium fluctuation theorems~\cite{jarzynski:prl:1997:neq-fe,crooks:pre:1999:cft,crooks:pre:2000:pathensemble}; these are particularly relevant to switching between multiple thermodynamic states.
Though the development of efficient estimators that utilize \emph{both} nonequilibrium switching trials and sampled equilibrium data generated in NCMC simulations remains an open challenge, it is at least straightforward to incorporate information from rejected NCMC proposals in the estimation of equilibrium averages~\cite{frenkel:2004:pnas:rejected-states,athenes-marinica:jcp:2010:free-energy-reconstruction}.



\begin{materials}

\section{WCA Dimer Simulations}
The dimer system considered here consists of two particles that interact via a double-well ``bonded'' potential in the interatomic distance~$r$,
\begin{eqnarray}
U_\mathrm{bond}(r) &=& h \left[ 1 - \frac{(r - r_0 - s)^2}{s^2}\right]^2
\end{eqnarray}
with $h = 5 \, k_\mathrm{B} T$, $r_0 = r_\mathrm{WCA}$, and $s = r_\mathrm{WCA}/2$, where $r_\mathrm{WCA} \equiv 2^{1/6} \sigma$.
Simulations denoted as ``vacuum'' contain only these two particles, while simulations denoted as ``solvated'' also interact with a dense bath of particles via the WCA nonbonded potential~\cite{weeks-chandler-andersen:1971:jcp:wca-fluid},
\begin{eqnarray}
U_\mathrm{WCA}(r) &=& \begin{cases}
4 \epsilon \left[\left(\frac{\sigma}{r}\right)^{12} - \left(\frac{\sigma}{r}\right)^6\right] + \epsilon , & r < r_\mathrm{WCA} \\
0 & r \ge r_\mathrm{WCA}
\end{cases},
\end{eqnarray}
with mass $m = 39.9$ amu, $\sigma = 3.4$ \AA, and $\epsilon = 120 \, k_\mathrm{B} T$.
The nonbonded WCA interaction is excluded between the two bonded particles.
The solvated system contains a total of 216 WCA particles at a reduced density of $\rho \sigma^3 = 0.96$.
For all simulations, the reduced temperature is $k_\mathrm{B} T / \epsilon = 0.824$.
A custom Python code making use of the GPU-accelerated OpenMM package~\cite{friedrichs:2009:j-comput-chem:openmm,eastman:2010:comp-sci-eng:openmm,eastman:2010:j-comput-chem:openmm} and the PyOpenMM Python wrapper~\cite{pyopenmm} was used to conduct the simulations.
All scripts are available for download from \url{http://simtk.org/home/ncmc}.

To ensure that observed differences were not due to changes in the stationary distribution of the integrator, we used generalized hybrid Monte Carlo (GHMC)~(\cite{gustafson:statistics-and-computing:1998:GHMC,jun-s-liu:MCMC,lelievre-stoltz-rousset:2010:free-energy-computations,lelievre-stoltz-rousset:2010:GHMC}) for all our simulations.
GHMC is based on a velocity Verlet discretization~\cite{swope:jcp:1982:velocity-verlet} of Langevin dynamics---the two are equivalent in the limit of small timesteps - but includes an acceptance/rejection step to correct for errors introduced by finite timesteps so that the stationary distribution is the exact equilibrium distribution.  We used a timestep of $0.002 \tau$, where $\tau = \sqrt{\sigma^2 m / \epsilon}$, and the collision rate was set to $\tau^{-1}$.
With this timestep, the acceptance probability is $99.929\pm0.001$\%; the resulting dynamics closely approximates Langevin dynamics.

In simulations employing instantaneous Monte Carlo moves, a perturbation $\Delta r$ to the interatomic distance $r$ of the dimer was chosen according to,
\begin{eqnarray}
\Delta r &=& \begin{cases}
+r_0 & \mathrm{if} \:\: r < 1.5 r_0 \\
-r_0 & \mathrm{if} \:\: 1.5 r_0 \le r \le 3 r_0 \\
0 & \mathrm{otherwise}
\end{cases} .
\end{eqnarray}
The dimer was contracted or expanded about the bond midpoint to generate a new configuration $x_\mathrm{new}$ with dimer extension $r_\mathrm{new}$ from the old configuration $x_\mathrm{old}$ with dimer extension $r_\mathrm{old}$, and the move accepted or rejected with the Metropolis-Hastings criterion~\cite{hastings:biometrika:1970:metropolis-hastings},
\begin{eqnarray}
A(x_\mathrm{new} | x_\mathrm{old}) &=& \mathrm{min}\left\{1, e^{-\beta[U(x_\mathrm{new})+U(x_\mathrm{old})]} J_r(x_\mathrm{old},x_\mathrm{new}) \right\} \nonumber \\
\end{eqnarray}
where the Jacobian ratio term $J_r(x_\mathrm{old}, x_\mathrm{new}) = (r_\mathrm{new} / r_\mathrm{old})^2$ accounts for the expansion and contraction of phase space due to the Monte Carlo proposals.

\color{black}
For simulations employing $T$-step NCMC moves, proposals were made by selecting a new velocity vector from the Maxwell-Boltzmann distribution, integrating $T$ steps of velocity Verlet dynamics~\cite{swope:jcp:1982:velocity-verlet} for all bath atoms as the dimer extension was driven from $r_\mathrm{old}$ to $r_\mathrm{new}$ in equal steps of size $\Delta r / T$, and accepting or rejecting based on the modified Metropolis criteria for symplectic integrators (Eqs.~\ref{equation:modified-metropolis-criteria} and \ref{equation:deterministic-acceptance-criteria}),
\begin{eqnarray}
A(X) &=& \mathrm{min}\left\{1, e^{-\beta[H(x_T)-H(x_0)]} J_r(x_0,x_T) \right\} .
\end{eqnarray}
The Jacobian ratio is also $(r_\mathrm{new}/r_\mathrm{old})^2$.
\color{black}
This scheme corresponds to a choice for the perturbation kernel of,
\begin{eqnarray}
\alpha_t(x,y) = \left[\frac{r(y)}{r(x)}\right]^2 \delta([r(y) - r(x)] - [\Delta r / T]) ,
\end{eqnarray}
where $r(x)$ denotes the dimer separation of configuration $x$.
The propagation kernel $K_t(x,y)$ corresponds to velocity Verlet dynamics where the dimer atoms are held fixed in space.
\color{black}
The final configuration after the MC or NCMC rejection procedure was stored and plotted to generate Fig.~\ref{figure:wca-trajectory}.

The mean acceptance probabilities for each switching time $\tau$ can be estimated via the sample mean
\begin{eqnarray}
\expect{A}_\tau &\approx& \frac{1}{N} \sum_{n=1}^N A(X_n)  .
\end{eqnarray}
For numerical stability, logarithms of $A(X_n)$ were stored, as $a_n \equiv \ln A(X_n)$. 
We then estimated $\ln \expect{A}_\tau$ (shown in Fig.~\ref{figure:wca-efficiency}) as
\begin{eqnarray}
\ln \expect{A}_\tau &\approx& - \ln N + \ln b + \sum_{n=1}^N e^{a_n - b} 
\end{eqnarray}
where $b \equiv \max_n a_n$.

Integrated autocorrelation times were estimated using the rapid scheme described in Section 5.2 of Ref.~\cite{chodera:jctc:2007:ptwham}.


The acceptance probabilities plotted in Fig.~\ref{figure:wca-efficiency} were estimated from a trajectory consisting of 10 000 iterations of 2048-step NCMC, with 500 steps of GHMC dynamics in between each NCMC trial, to ensure equilibrium sampling.
Prior to each 2048-step NCMC iteration, a velocity from the Maxwell-Boltzmann distribution was selected, and NCMC trial moves with varying switching times were conducted solely to accumulate statistics.
The statistical error in the estimate of $\ln \expect{A}_\tau$ and the computed relative efficiency over instantaneous Monte Carlo was estimated by 1000 bootstrap trials, in which the dataset of 10 000 work samples was resampled with replacement in each bootstrap trial and 95\% confidence intervals computed from the distribution over bootstrap replicates.

\begin{figure}[tbp]
\noindent
\resizebox{\columnwidth}{!}{\includegraphics{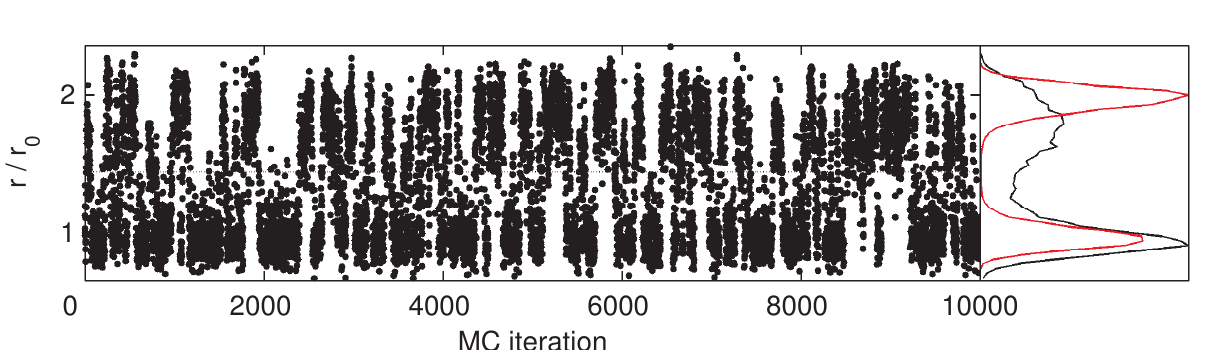}}
\caption{{\bf Umbrella sampling simulation of the dimer in WCA solvent.}
\emph{Left:} The dimer extension $r$ as a function of simulation iteration.
\emph{Right:} The histogram accumulated over the trajectory, with the observed histogram in black and the reweighted histogram (corrected for the applied umbrella bias potential) in red.
\label{figure:wca-dimer-umbrella}}
\end{figure}

The reference distribution for the interparticle distribution $\mathcal{P}(r)$ plotted in red on the right side of Fig.~\ref{figure:wca-trajectory} was computed analytically for the vacuum system,
\begin{eqnarray}
\mathcal{P}_\mathrm{vac}(r) &\propto& 4 \pi r^2 e^{-\beta U_\mathrm{bond}(r)} .
\end{eqnarray}
For the solvated system, this distribution was estimated from an umbrella sampling simulation employing a modified bonded potential intended to remove the barrier in between compact and extended states, 
\begin{eqnarray}
U_\mathrm{umbrella}(r) &=& k_\mathrm{B} T \ln r^2 + \theta(r_\mathrm{min} - r) (K/2) [r-r_\mathrm{min}]^2 \nonumber \\
&& \mbox{} + \theta(r - r_\mathrm{max}) (K/2) [r-r_\mathrm{max}]^2 ,
\end{eqnarray}
where $r_\mathrm{min} = r_0$, $r_\mathrm{max} = 2.05 r_0$, and $K = k_\mathrm{B} T / \eta^2$, with $\eta = 0.3$ \AA, and $\theta(r)$ is the Heaviside function that assumes a value of unity for $r \ge 0$ and zero otherwise.
The true solvated interparticle distribution $p(r)$ was estimated by reweighting the data produced from this simulation, using the relationship
\begin{equation}
\mathcal{P}_\mathrm{sol}(r) \propto \frac{\sum_{n=1}^N \delta(r - r_n) \, e^{-\beta [U_\mathrm{bond}(r_n) - U_\mathrm{umbrella}(r_n)]}}{\sum_{n=1}^N e^{-\beta [U_\mathrm{bond}(r_n) - U_\mathrm{umbrella}(r_n)]}}
\end{equation}
where $r_n$ denotes the bond separation for sample $n$, and a finite-width histogram bin was used instead of the delta function $\delta(r)$.
\end{materials}


\begin{acknowledgments}
The authors thank Gabriel Stoltz (CERMICS, \'{E}cole des Ponts ParisTech); Jed W.~Pitera (IBM Almaden Research Center); {\color{black}Manuel Athen\`{e}s (CEA Saclay)}; {\color{black}Firas Hamze (D-Wave Systems)}; Yael Elmatad, Anna Schneider, Paul Nerenberg, Todd Gingrich, David Chandler, David Sivak, Phillip Geissler, Michael Gr\"{u}nwald, and Ulf R{\"o}rb{\ae}ck Pederson (University of California, Berkeley); Vijay S. Pande (Stanford University); and Suriyanarayanan Vaikuntanathan, Andrew J.~Ballard, and Christopher Jarzynski (University of Maryland), and Huafeng Xu (D.~E.~Shaw Research) for enlightening discussions on this topic and constructive feedback on this manuscript, {\color{black}as well as the two anonymous referees for their helpful suggestions for improving clarity.}
JPN was supported by the U.S.~Department of Energy by Lawrence Livermore National Laboratory under Contract DE-AC52-07NA27344.
GEC was funded by the Helios Solar Energy Research Center, which is supported by the Director, Office of Science, Office of Basic Energy Sciences of the U.S.~Department of Energy under Contract No.~DE-AC02-05CH11231.
DDLM was funded by a Director's Postdoctoral Fellowship from Argonne National Laboratory.
JDC was supported through a distinguished postdoctoral fellowship from the California Institute for Quantitative Biosciences (QB3) at the University of California, Berkeley.
Additionally, the authors are grateful to {\sc OpenMM} developers Peter Eastman, Mark Friedrichs, Randy Radmer, and Christopher Bruns for their generous help with the {\sc OpenMM} GPU-accelerated computing platform and associated {\sc PyOpenMM} Python wrappers.
This research was performed under the auspices of the U.S. Department of Energy by Lawrence Livermore National Laboratory under Contract DE-AC52-07NA27344 and by UChicago Argonne, LLC, Operator of Argonne National Laboratory (Argonne).  Argonne, a U.S. Department of Energy Office of Science laboratory, is operated under Contract No. DE-AC02-06CH11357.
The U.S. Government retains for itself, and others acting on its behalf, a paid-up nonexclusive, irrevocable worldwide license in said article to reproduce, prepare derivative works, distribute copies to the public, and perform publicly and display publicly, by or on behalf of the government.
\end{acknowledgments}



\end{article}


\eject
\appendix


\section{Proof that NCMC preserves the equilibrium distribution}
\label{appendix:NCMC-proof}

Following the proof for GHMC in Ref.~\cite{lelievre-stoltz-rousset:2010:free-energy-computations}, here we show that NCMC preserves the equilibrium distribution.  The expected acceptance rate for NCMC moves initiated from $(x,\lambda)$ is,
\begin{eqnarray}
\alpha(x,\lambda) &\equiv& \int d\Lambda \int dX \, \rho(X, \Lambda | x, \lambda) \, A(X|\Lambda) .
\end{eqnarray}

Suppose that we have a variate $(x^{(n)},\lambda^{(n)})$ drawn from the equilibrium distribution $\pi(x,\lambda)$.
The probability density of the next value in the chain, $p(x^{(n+1)},\lambda^{(n+1)})$, has contributions from two scenarios: when the candidate variate is rejected and when it is accepted.
The contribution from rejecting the candidate and flipping the momentum such that $(x^{(n+1)},\lambda^{(n+1)}) = (\tilde{x}^{(n)},\lambda^{(n)})$ is,
\begin{eqnarray}
\int dx \sum_\lambda \pi(x,\lambda) [1 - \alpha(x,\lambda)] \delta(\tilde{x}-x^{(n+1)}) \delta(\lambda - \lambda^{(n+1)}) & = & \pi(\tilde{x}^{(n+1)}, \lambda^{(n+1)}) [1 - \alpha(\tilde{x}^{(n+1)},\lambda^{(n+1)})] .
\label{eq:rejection-term}
\end{eqnarray}
The latter contribution from accepting the candidate such that $(x^{(n+1)},\lambda^{(n+1)}) = (x_T,\lambda_T)$ is,
\begin{eqnarray}
& & \int dx \sum_{\lambda}  \pi(x,\lambda) \int dX \int d\Lambda \, \rho(X,\Lambda | x, \lambda) \, A(X|\Lambda) \delta(x_T-x^{(n+1)}) \delta(\lambda_T - \lambda^{(n+1)}) \nonumber \\
& = & \int dx_0 \sum_{\lambda_0} \int dX \int d\Lambda \, \left[ \pi(x_0,\lambda_0) \, \rho(X,\Lambda | x_0, \lambda_0) \, A(X|\Lambda)  \right] \delta(x_T-x^{(n+1)}) \delta(\lambda_T - \lambda^{(n+1)}) \nonumber \\
& = & \int dx_T \sum_{\lambda_T} \int d\tilde{X} \int d\tilde{\Lambda} \, \left[ \pi(\tilde{x}_T,\lambda_T) \, \rho(\tilde{X},\tilde{\Lambda} | \tilde{x}_T, \lambda_T) \, A(\tilde{X}|\tilde{\Lambda})  \right] \delta(x_T-x^{(n+1)}) \delta(\lambda_T - \lambda^{(n+1)}) \nonumber \\
& = & \pi(\tilde{x}^{(n+1)},\lambda^{(n+1)}) \alpha(\tilde{x}^{(n+1)}, \lambda^{(n+1)}),
\label{eq:acceptance-term}
\end{eqnarray}
\color{black}
where $\rho(X,\Lambda | x_0, \lambda_0) \equiv  \Pi(X | x_0, \Lambda) P(\Lambda | x_0, \lambda_0)$ is the probability of generating the trajectory-protocol pair $(X,\Lambda)$ from $(x_0,\lambda_0)$, and the pathwise detailed balance condition (Eq. \ref{equation:pathwise-detailed-balance}) is used to produce the quantity in brackets.
\color{black}

Taking the sum of Eqs. \ref{eq:rejection-term} and \ref{eq:acceptance-term}, we find that the equilibrium distribution is preserved,
\begin{eqnarray}
p(x^{(n+1)},\lambda^{(n+1)}) = \pi(x^{(n+1)},\lambda^{(n+1)})
\end{eqnarray}

By analogous reasoning, maintaining the momentum upon rejection, $(x^{(n+1)},\lambda^{(n+1)} = (x^{(n)},\lambda^{(n)})$, and flipping it upon acceptance, $(x^{(n+1)},\lambda^{(n+1)} = (\tilde{x}_T,\lambda_T)$ will also preserve the equilibrium distribution.



\section{Acceptance criteria for overdamped Langevin (Brownian) integrator of Ermak and Yeh}

A common integrator for Brownian dynamics (the overdamped regime of Langevin dynamics), in which only coordinates $x$ are explicitly integrated, is given by Ermak and Yeh~\cite{ermak-yeh:cpl:1974:brownian-dynamics,ermak:jcp:1975:brownian-dynamics}.
\color{black}In our notation, where the perturbed coordinates $x_t^*$ are propagated by one step of the stochastic integrator to obtain $x_t$, application of the propagation kernel $K(x_t^*, x_t)$ can be written,
\begin{eqnarray}
x_t &=& x_t^* - \frac{\Delta t}{\gamma m} F_t(x_t^*) + \sqrt{2} \left( \frac{\Delta t}{\gamma m}\right)^{1/2} \,\xi_t
\end{eqnarray}
\color{black}
where $m$ is the particle mass, $F_t(x) \equiv -(\partial/\partial x) H_t(x)$ is the (potentially time-dependent) systematic force, and $\gamma$ is an effective collision frequency or friction coefficient with units of inverse time.
The noise history $\xi_t$ for each degree of freedom is a normal random variate with zero mean and variance $\beta^{-1}$, drawn from the distribution
\begin{eqnarray}
\phi(\xi_t) &=& \frac{1}{\sqrt{2 \pi \beta^{-1}}} \exp\left[-\frac{\beta}{2} \xi_t^2\right]
\end{eqnarray}

In NCMC, every application of the propagation kernel $K_t$ produces a transition $x_t^* \rightarrow x_t$ determined by the noise history variable $\xi_t$, there is a corresponding $\tilde{\xi}_t$ that generates the opposite step, $x_t \rightarrow x_t^*$.
By noting
\color{black}
\begin{eqnarray}
x_t  &=& x_t^* - \frac{\Delta t}{\gamma m} F_t(x_t^*) + \sqrt{2} \left( \frac{\Delta t}{\gamma m}\right)^{1/2} \,\xi_t \nonumber \\
x_t^* &=& x_t - \frac{\Delta t}{\gamma m} F_t(x_t) + \sqrt{2} \left( \frac{\Delta t}{\gamma m}\right)^{1/2} \,\tilde{\xi}_t
\end{eqnarray}
\color{black}
we can compute the relationship,
\begin{eqnarray}
\tilde{\xi}_t &=&  \frac{1}{\sqrt{2}} \left(\frac{\Delta t}{\gamma m}\right)^{1/2} \left[F_t(x_t) + F_t(x_t^*)\right] - \xi_t,
\end{eqnarray}

Then, the ratio of transition kernels appearing in Eq.~\ref{equation:ratio-of-propagation-kernels} can be written in terms of noise history $\xi_t$ and the computed reverse noise history $\xi_t^*$,
\begin{eqnarray}
\Delta \mathcal{S}(X) = - \ln \prod_{t=1}^T \frac{K_t(x_t, x_t^*)}{K_t(x_t^*, x_t)}  = - \ln \prod_{t=1}^T \frac{\phi(\tilde{\xi}_t) \left|\frac{\partial x_t^*}{\partial \tilde{\xi}_t}\right|}{\phi(\xi_t) \left|\frac{\partial x_t}{\partial \xi_t}\right|} = - \ln \prod_{t=1}^T \frac{\phi(\tilde{\xi}_t)}{\phi(\xi_t)} = - \ln \prod_{t=1}^T \exp\left[ -\frac{\beta}{2} ({\tilde{\xi}_t}^2 - {\xi_t}^2)\right] = \frac{\beta}{2} \sum_{t=1}^T ({\tilde{\xi}_t}^2 - {\xi_t}^2)
\label{equation:Ermak-action}
\end{eqnarray}
where the tildes are dropped because the microstate $x$ contains no momenta.
The quantity $\left|\partial x_t/\partial \xi_t\right|$ represents the Jacobian for the change of variables from the $\xi_t$ to $x_t$, and the Jacobians in the numerator and denominator cancel.
The quantity in Eq. \ref{equation:Ermak-action} can easily be computed during integration.


\section{Acceptance criteria for Langevin integrator of Brooks, Br\"{u}nger, and Karplus (BBK)}

The Br\"{u}nger-Brooks-Karplus (BBK) stochastic integrator~\cite{brunger-brooks-karplus:cpl:1984:bbk-integrator,pastor-brooks-szabo:mol-phys:1988:bbk-integrator} is a popular integrator for simulating Langevin dynamics.
\color{black}In our notation, where the perturbed coordinates $x_t^*$ are propagated by one step of the stochastic integrator to obtain $x_t$, application of the propagation kernel $K(x_t^*, x_t)$ can be written,
\begin{eqnarray}
v'_{t} &=& v^*_t + \frac{\Delta t}{2 m} \left(F_t(r^*_t) - \gamma m v^*_t + \sqrt{\frac{2 \gamma m}{\Delta t}} \xi_t\right) \nonumber \\
r_{t} &=& r^*_t + \Delta t \, v'_t \nonumber \\
v_{t} &=& \frac{1}{1 + \frac{\gamma \Delta t}{2}} \left[ v'_{t} + \frac{\Delta t}{2 m} \left( F_t(r_t) + \sqrt{\frac{2 \gamma m}{\Delta t}} \xi'_{t} \right) \right]
\end{eqnarray}
where we have used a velocity Verlet discretization of the BBK integrator~\cite{schlick,lelievre-stoltz-rousset:2010:free-energy-computations}.
\color{black}
Here $r_t$ and $v_t$ denote the respective Cartesian position and velocity components of the microstate $x_t$,  $\gamma$ the effective collision frequency with units of inverse time, and $m$ the particle mass.  
$v'_t$ is an auxiliary variable used only in simplifying the mathematical representation of the integration scheme.
Note that we require \emph{two} random variates, $\xi_t$ and $\xi'_t$, per degree of freedom per timestep in order for this scheme to be able to generate both the forward trajectory $X$ and its time-reverse $\tilde{X}$ (see, e.g., Section 2.2.3.2 of \cite{lelievre-stoltz-rousset:2010:free-energy-computations}).

The noise history terms $\xi_t$ and $\xi'_t$ are normal random variates with zero mean and variance $\beta^{-1}$.
Their joint distribution can therefore be written,
\begin{eqnarray}
\psi(\xi_t,\xi'_t) &=& \frac{1}{2 \pi \beta^{-1}} \exp\left[-\frac{\beta}{2} \left(\xi_t^2 + {\xi'_t}^2\right) \right] .
\end{eqnarray}

For every step $x_t^* \rightarrow x_t$, the positions and velocities undergo a transition $(r_t^*,v_t^*) \rightarrow (r_t,v_t)$ determined by the noise variables $(\xi_t, \xi'_t)$.
A corresponding choice of noise variables $(\tilde{\xi}_t,\tilde{\xi}'_t)$ will generate the reverse step, $\tilde{x}_t \rightarrow \tilde{x}_t^*$, carrying $(r_t,-v_t) \rightarrow (r_t^*,-v_t^*)$
With a little algebra, it is seen that,
\begin{eqnarray}
\tilde{\xi}_t &=& \xi'_t - \sqrt{2 \gamma \, m \, \Delta t} \, v_t \nonumber \\
\tilde{\xi}'_t &=& \xi_t - \sqrt{2 \gamma \, m \, \Delta t} \, v_t^* .
\end{eqnarray}
In order to write the ratio of transition kernels appearing in Eq.~\ref{equation:ratio-of-propagation-kernels} in terms of noise variables $(\xi_t,\xi'_t)$ and the computed reverse noise variables $(\tilde{\xi}_t, \tilde{\xi}'_t)$, we must first compute the Jacobian $J(\xi_t, \xi'_t)$ because the random variates are not in Cartesian space,
\begin{eqnarray}
J(\xi_t, \xi'_t) &\equiv& \det \left[ 
\begin{array}{cc}
\frac{\partial r_t}{\partial \xi_t} & \frac{\partial v_t}{\partial \xi_t} \\
\frac{\partial r_t}{\partial \xi'_t} & \frac{\partial v_t}{\partial \xi'_t}
\end{array}
\right] ,
\end{eqnarray}
which can be shown to be independent of $\xi_t$ and $\xi'_t$.
The conditional path action difference can now be computed,
\begin{eqnarray}
\Delta \mathcal{S}(X) = - \ln \prod_{t=1}^T \frac{K_t(\tilde{x}_t, \tilde{x}_t^*)}{K_t(x_t^*, x_t)}  = - \ln \prod_{t=1}^T \frac{\psi(\tilde{\xi}_t,\tilde{\xi}'_t) J(\tilde{\xi}_t,\tilde{\xi}'_t)}{\psi(\xi_t,\xi'_t) J(\xi_t,\xi'_t)}  = \frac{\beta}{2} \sum_{t=1}^T \left[ \left(\left.{\tilde{\xi}_t}\right.^2 + \left.{\tilde{\xi}'_t}\right.^2\right) - \left({\xi_t}^2 + {\xi'_t}^2\right)\right]
\end{eqnarray}
where the ratio of Jacobians $J(\tilde{\xi}_t, \tilde{\xi}'_t) / J(\xi_t, \xi'_t)$ cancels because the Jacobians are independent of the noise variates.


\color{black}
\section{Derivation of effective correlation time for mixed MD/NCMC sampling}

For simplicity, we make the assumption that the system of interest has two long-lived conformational states of equal population with dimer extensions $r_c$ and $r_e$.
This assumption holds to good approximation for the WCA dimer example considered here, and may apply to other systems of interest as well.
We assume that the correlation time for a fixed number of MD simulation steps per iteration is given by $\tau_\mathrm{MD}$, and describe the probability of finding the system ends up in a given conformational state after one iteration by a $2 \times 2$ column-stochastic transition matrix ${\bf T}_\mathrm{MD}$,
\begin{eqnarray}
{\bf T}_\mathrm{MD} &=& \left[
\begin{array}{cc}
1-\alpha & \alpha \\
\alpha & 1 - \alpha
\end{array}
\right]
\end{eqnarray}

For a $2 \times 2$ system whose time evolution is governed by the column stochastic transition matrix ${\bf T}$, we can write the autocorrelation function for the dimer extension $r$ as
\begin{eqnarray}
C(n \Delta t) &=& \expect{r(0) \, r(n \Delta t)} \nonumber \\
&=& \left[ \begin{array}{cc} r_c & r_e \end{array} \right] {\bf T}^n \left[ \begin{array}{cc} 1/2 & 0 \\ 0 & 1/2 \end{array} \right] \left[ \begin{array}{c} r_c \\ r_e \end{array}\right] \nonumber \\
&=& \left[ \begin{array}{cc} r_c & r_e \end{array} \right] {\bf U} \left[ \begin{array}{cc} 1 & 0 \\ 0 & \mu^n \end{array} \right] {\bf U}^{-1}  \left[ \begin{array}{cc} 1/2 & 0 \\ 0 & 1/2 \end{array} \right] \left[ \begin{array}{c} r_c \\ r_e \end{array}\right] \nonumber \\
&=& (C_0 - C_\infty) \mu^n + C_\infty
\end{eqnarray}
where the transition matrix ${\bf T}$ has unitary eigenvalue decomposition ${\bf U} {\bf M} {\bf U}^{-1}$, and $\mu$ is the non-unit eigenvalue of ${\bf T}$.
The constants are $C_0 = (1/2) (r_c^2 + r_e^2)$ and $C_\infty = (1/4) (r_c + r_e)^2$.

Relating this to the autocorrelation time $\tau$ estimated from a timeseries, intended to reflect the fit to
\begin{eqnarray}
C(t) = (C_0 - C_\infty) e^{-t/\tau} + C_\infty
\end{eqnarray}
we can see that $\tau = - 1 / \ln \mu$.
We then determine that the correlation time $\tau_\mathrm{MD} = -1/\ln \mu_\mathrm{MD}$, with $\mu_\mathrm{MD} = 1 - 2 \alpha$.

Similarly, we can write the probability that the NCMC step will carry the system from one conformational state to another in terms of the acceptance probability $\gamma$, which we assume to be symmetric,
\begin{eqnarray}
{\bf T}_\mathrm{NCMC} &=& \left[
\begin{array}{cc}
1-\gamma & \gamma \\
\gamma & 1 - \gamma
\end{array}
\right]
\end{eqnarray}
where we have correlation time $\tau_\mathrm{NCMC} = -1/\ln \mu_\mathrm{NCMC}$ and $\mu_\mathrm{NCMC} = 1 - 2 \gamma$.

The effective transition matrix ${\bf T}_\mathrm{eff}$ for iterations consisting of MD simulation steps followed by an NCMC trial is given by
\begin{eqnarray}
{\bf T}_\mathrm{eff} = {\bf T}_\mathrm{MD} {\bf T}_\mathrm{NCMC} &=&  \left[
\begin{array}{cc}
1-\alpha & \alpha \\
\alpha & 1 - \alpha
\end{array}
\right] \left[
\begin{array}{cc}
1-\gamma & \gamma \\
\gamma & 1 - \gamma
\end{array}
\right] 
= \left[ \begin{array}{cc}
1 - (\alpha + \gamma) & (\alpha + \gamma) - 2 \alpha \gamma \\
(\alpha + \gamma) - 2 \alpha \gamma & 1 - (\alpha + \gamma)
\end{array} 
\right]
\end{eqnarray}
where the effective correlation time $\tau_\mathrm{eff} = -1/\ln \mu_\mathrm{eff}$, with $\mu_\mathrm{eff} = 1 - 2[(\alpha + \gamma) - 2 \alpha \gamma]$.
Substituting in $\alpha = (1 - e^{-1/\tau_\mathrm{MD}})/2$ and $\gamma = (1 - e^{-1/\tau_\mathrm{NCMC}})/2$, we obtain
\begin{eqnarray}
\tau_\mathrm{eff} &=& - \frac{1}{\ln\left[ 1 - (1 - e^{-1/\tau_\mathrm{MD}}) - (1 - e^{-1/\tau_\mathrm{NCMC}}) + (1 - e^{-1/\tau_\mathrm{MD}})(1 - e^{-1/\tau_\mathrm{NCMC}})\right]} \nonumber \\
&=&  - \frac{1}{\ln\left[e^{-1/\tau_\mathrm{MD}} e^{-1/\tau_\mathrm{NCMC}}\right]} =  \frac{1}{\tau_\mathrm{MD}^{-1} + \tau_\mathrm{NCMC}^{-1}} = \frac{\tau_\mathrm{MD} \, \tau_\mathrm{NCMC}}{\tau_\mathrm{MD} + \tau_\mathrm{NCMC}} \label{equation:effective-correlation-time}
\end{eqnarray}
As a check of the accuracy of Eq.~\ref{equation:effective-correlation-time}, we note that for MD with 2048-step NCMC switching, we compute $\tau_\mathrm{eff} \approx 4.0$ iterations, using only $\tau_\mathrm{MD} = 299.8$ iterations and the NCMC acceptance probability $\gamma = 12.1\%$. The actual correlation time measured from a 10 000 iteration simulation is computed to be $\tau_\mathrm{eff} = 4.0$.

\color{black}

\end{document}